\documentclass[prb,preprint,showpacs]{revtex4}
\usepackage{graphicx}
\usepackage[T1]{fontenc}
\usepackage{amssymb,amsmath}
\begin{document}
\title{Influence of the microstructure on the magnetism of Co-doped ZnO\ thin films.}
\author{A. Fouchet} 
\affiliation{Laboratoire CRISMAT, UMR CNRS--ENSICAEN(ISMRA) 6508, 
             6, Bld. du Mar\'echal Juin, F-14050 Caen, France}
\author{W. Prellier\thanks{%
prellier@ensicaen.fr}} 
\affiliation{Laboratoire CRISMAT, UMR CNRS--ENSICAEN(ISMRA) 6508, 
             6, Bld. du Mar\'echal Juin, F-14050 Caen, France}
\author{B. Mercey}
\affiliation{Laboratoire CRISMAT, UMR CNRS--ENSICAEN(ISMRA) 6508, 
             6, Bld. du Mar\'echal Juin, F-14050 Caen, France}
\date{\today}

\begin{abstract}
The prediction of ferromagnetism at room temperature in Co-ZnO thin films
has generated a large interest in the community due to the possible
applications. However, the results are controversial, going from
ferromagnetism to non-ferromagnetism, leading to a large debate about its
origin (secondary phase, Co clusters or not). By carefully studying the
micro-structure of various Co-ZnO films, we show that the Co$^{2+}$ partly
substitutes the ZnO\ wurtzite matrix without forming Co clusters.\
Surprisingly, the ferromagnetism nature of the films disappears as the Co
content increases{\bf .} In addition, our results suggest that the observed
ferromagnetism is likely associated to a large amount of defects- close to
the interface and strongly depending on the growth temperature- which may
explained the spreading of the results.
\end{abstract}
\maketitle

\newpage

\section{Introduction}

Recently, Diluted Magnetic Semiconductors (DMS)\cite{DMS1,DMS2,DMS3,DMS3b}
have become a very attractive subject due to the possibility of room
temperature ferromagnetism in wide--band gap oxides.\cite{WidBand} Such
materials might be integrated into semiconductors devices opening the route
to spin-electronic technology at high temperature. First reports were
dedicated to the cobalt-doped ZnO,\cite{CoZnO2} and TiO$_2$,\cite
{DMS2,CoTiO1} and the manganese-doped ZnO.\cite{MnZnO}\ Following these
results, many materials showing ferromagnetism have been isolated so far.%
\cite{F,F5,F15,F16}\ Surprisingly, other reports did not evidence any
ferromagnetism in the compounds \cite{NoFM,NoFM1,NoFM2,NoFM7} especially
when made in bulk materials,\cite{NoFM3} which has led to a hot debate about
the origin of ferromagnetism in these materials.\ In particular, it is not
clear whether ferromagnetism is originated from clusters,\cite
{Clusters,clusters1,clusters2} secondary phases \cite{Second} or \smallskip
it is an intrinsic phenomenon.\cite{Int2a,Int1} Thus, it is important to
answer the following questions:

- why are there so many non-reproducible reports?

- why is ferromagnetism relatively independent of the dopant and its
concentration?\cite{Int1}

- is the ferromagnetism intrinsic in Co-doped ZnO\ films?

During the last 5 years, several theoretical predictions raised the
possibility of ferromagnetism with Curie temperature (T$_C$) above room
temperature in $3d$-transition-metal-doped ZnO. Dietl {\it et al.}\cite{DMS1}
suggested that wide band gap semiconductors are candidates for a high T$_C$
and a large magnetization when 5\% Mn is substituted into $p$-type [10$^{20}$%
cm$^{-3}$]. Using $ab$-initio band structure calculation, Sato {\it et al.}%
\cite{DMS4} predicted a stabilization of the ferromagnetic state in $3d$%
-transition metal doped ZnO. Recently, Coey {\it et al.}\cite{Int1} proposed
a model for high temperature ferromagnetism in dilute $n$-type magnetic
semiconductor. This model is based on the formation of bound polaron
magnetic mediated by shallow donor electrons.

>From these different theories and the experimental results, it appears that
the role of defects is important.\cite{Bouzenar} Different scenarios are
thusly possible.\ First, the presence of oxygen vacancies or interstitial
zinc possibly result in an increase of the conductivity since they may
create donor electrons.\cite{Int1} Second, a large amount of cationic defect
or excess of dopant can also lead to cluster or secondary phases. In
addition, the defects might come from the growth techniques and the
conditions used (temperature, oxygen pressure).\cite{DMS3b} The spreading of
the results by the different groups probably comes from the strong influence
of the deposition parameters on the structure which is correlated to the
magnetic properties.

In order to elucidate the origin of ferromagnetism and answer some of these
questions, we have first undertaken a detailed microstructural analysis
using transmission electron microscopy on the host matrix, ZnO, grown at
various temperature. Second, we have correlated the magnetic properties and
the microstructure of a series of cobalt-doped ZnO films.\ Third, we find
the experimental tendency that the presence of defects in Co-ZnO\ films is
necessary for the observation of room temperature ferromagnetism. Finally,
we turn to an interpretation based on theoretical model.

\section{Experimental}

The ZnO and Co-ZnO films were grown utilizing the pulsed laser deposition
technique (Lambda Physik, KrF laser $\lambda =248$ $nm$)\cite{PLD} by firing
the laser alternatively on a zinc metal target (99.995 $\%$) and on a cobalt
metal target (99.995 $\%$). These target were purchased (NEYCO, France) and
were used without further preparation. All films are deposited on
(0001)-oriented Al$_2$O$_3$ substrates at a constant temperature under a
flux of pure oxygen gas. Typical thickness of the films is about 300nm.The
structural study was done by X-Ray diffraction (XRD) using a Seifert XRD
3000P for the $\Theta -2\Theta $ scans and the $\omega $-scan (tilting) with
(Cu, K$\alpha 1$ radiation $\lambda =0.15406$ $nm$). The electron
diffraction (ED) and electron microscopy observations were performed on JEOL
2011FEG electron microscope (tilt $\pm 45{{}^{\circ }}$) equipped with
Energy Dispersive Spectrometer (EDS) analyzer. This microscope is also
fitted with a Scanning Transmission Electron Microscope equipment (STEM) to
carry out high resolution scanning electron microscopy or EDS cartography.
The resistivity measurements were measured in a Physical Properties
Measurements System (PPMS) Quantum Design. Silver electrodes were deposited
by thermal evaporation through a mask and contact between thin film and
sample holder were realized with ultrasonic bonding (wire of Al-Si 99/1).
The magnetization measurements were collected using a superconducting
quantum interference device based magnetometer SQUID (Quantum Design MPMS-5).

\section{Results}

\subsection{Temperature-dependance of the microstructure of ZnO\ films}

Before understanding the magnetic properties of the cobalt-doped ZnO, a
study of the influence of the deposition parameters on microstructure of the
host matrix, ZnO, was performed. For this, the structural quality of a
series of ZnO films grown under different growth conditions was
investigated. One of the main deposition parameter is the substrate
temperature (which is known to be of the most important parameter)\cite
{JAPArnaud} and the films were investigated using the XRD. Out-of-plane
lattice parameter close 0.518 nm\ was obtained in the temperature range
500-700 ${{}^{\circ }}C$, in agreement with the expected one.\cite{bulkvalue}
In addition, the \smallskip rocking curve recorded around the (0002)
reflection measured for the films grown at 500, 600 and 700 ${{}^{\circ }}C$
are 0.4, 0.25 and 0.32${^{\circ }}${,\ respectively. Such values are
commonly obtained for an oxide thin film and reflect, at least from these
measurements, an average good crystallinity of the overall film (the
instrument resolution is close to 0.2}${{}^\circ}$){. }This means that,
using a routine XRD analysis, there is no noticeable change and indicates
that the subtle differences can only be seen with deeper analysis such as
asymmetrical XRD or optical measurements.\cite{JAPArnaud} To complete this
analysis at such atomic scale, a microstructural analysis using transmission
electron microscopy (TEM) was undertaken.

Figure 1 (a) and (b) show typical TEM images of two films grown at 600 and
700${{}^{\circ }}C$, respectively.\ The corresponding selected area
diffraction are also presented.\ The images show a good crystallinity of the
films. More precisely, the electron diffraction patterns (inset of Fig.1a,
1b) reveal two interesting features.\ First, the ED patterns reveal 6 spots
as expected from a wurtzite structure (in agreement with the XRD)\cite
{JAPArnaud} and a film oriented with the $c$-axis perpendicular to the plane
of the substrate. Second, the spots are well defined for the film at 700 ${%
{}^{\circ }}C$ but some broadening or parasite spots can be observed for the
film grown at 600 ${{}^{\circ }}C$. This suggests that the film is less
crystallized when the temperature of the substrate is decreased,\ in
agreement with our previous reports.\cite{JAPArnaud} Note that the decrease
of the crystalline quality with temperature is confirmed by the analysis of
the films deposited at a temperature lower than 600 ${{}^{\circ }}C$ (not
shown).\ 

Furthermore, Fig.1c shows a plane view of the interfacial area between the
film and the substrate of a film grown at 700 ${{}^{\circ }}C$.\ In this
picture a ''moir\'{e}'' contrast, resulting from the superimposition of the
two lattices, is observed. The Fourier transform from this part of the image
(insert of Fig.1c) clearly shows that the two lattices are slightly
misoriented (about 1${{}^{\circ }}$). However this result raises the
understanding of the misorientation of 30${{}^{\circ }}$ between the film
and the substrate already observed by the XRD.\cite{JAPArno1} In order to
explain such a discrepancy, two features can be put forward.\ Firstly, this
picture shows only the early stage of the growth, i.e. the first few cells
grew on top of the substrate. The array of the oxygen atoms of the sapphire
substrate, provides a good template to accommodate the first zinc ad-atoms
despite the large difference between the lattice parameters of the two
structures (lattice mismatch 16 \%).\cite{LMismatch}\ Secondly, when the
thicker layers are grown, the lattice mismatch between the two structures
plays the most important role and the cell of the film will rotate in order
to accommodate this large mismatch. The thickness of this ''accommodation
layer'' depends on the temperature of the growth.\ In others words, this
''accommodation layer'' is thin when the temperature is high whereas the
''accommodation layer'' is thick at a lower temperature. Such result is in
agreement with Ashkenov {\it et al}. where a very thin nucleation layer (2-3
monolayers) with no cracks is observed on top of the Al$_2$O$_3$ substrate
for a ZnO film grown at high temperature (800 ${{}^{\circ }}C$).\cite
{nucleation layer}

Since the crystallinity of the film grown at 700 ${{}^{\circ }}C$ is higher
than the one at 600 ${{}^{\circ }}C$, we have chosen to look only at a less
crystallized film (i.e. 600 ${{}^{\circ }}C$) and its evolution along the $c$%
-axis direction. Different areas of a film grown at 600 ${{}^{\circ }}C$
were thusly observed. Figure 2 shows the electron diffraction patterns of
this film, close to the surface (Fig.2 (a)) and close to the interface
(Fig.2 (d)) and in between (Fig.2 (b) and (c)). When considering at the
evolution of the ED patterns, it is clear that the 6 well defined peaks,
close to the surface becomes a series of two rings, indicating a poor
crystallization close to the interface. These diffraction images show that
the film shows a gradient of crystallinity from the interface to the surface
which is mostly due to the important lattice mismatch between the substrate
and the film. However, we have noticed that such effect is enhanced when
decreasing the temperature of the substrate. In other words, the less
crystallized layer of the film close to the interface is increased when the
temperature of deposition is decreased. It is also important to note that
the polycrystalline layers close to the interface can not be observed with
the standard XRD characterizations. Thus, it is necessary to perform
microstructural measurements because it gives a local characterization
whereas the XRD measurements give only an idea of the average crystallinity
of the film. To explain the divergence of the 30${{}^{\circ }}$ rotation
observed by XRD measurements between the two lattices and the ''moir\'{e}''
with a tilting of 1${{}^{\circ }}$ at the interface observed by HRTEM, the
lattice mismatch must also be taken into account.

To summarize, the large strain induces the formation of a buffer layer at
the interface, which can be polycrystalline.\ However, it can be reduced by
an increasing of the deposition temperature (see above). This means that a
stabilization of the structure with a rotation of the film parameters of 30${%
{}^{\circ }}$ compared to the substrate is favoring by the increasing of the
substrate temperature. Moreover, the use of high temperature increases the
mobility of the ad-atoms and the size of the ZnO crystallites.\cite
{JAPArnaud} We believe that this is an explanation of the difference
observed in the literature of Co-doped ZnO films as detailed hereafter. To
confirm this, we have undertaken similar microstructural study of Co-doped
ZnO film, following the same approach.

\subsection{Temperature-dependance of the microstructure of Co-doped ZnO\
films}

\subsubsection{The case of 1.66 \%\ Co}

A low cobalt concentration (1.66 \%) has been chosen in agreement with
previous results because for such a composition, the presence of a secondary
phase in the film is unlikely. The dependance of the magnetic properties on
the growth conditions of a Co-ZnO\ film was analyzed.\ The films were grown
at different deposition temperatures (500, 600 and 700 ${{}^{\circ }}C$) and
the pressure of oxygen was varying from 0.05 to 0.15 $Torr$. As seen
previously, the temperature tunes the crystallinity of the film. The oxygen
pressure influences the resistivity with the creation of oxygen
non-stoichiometry or intersticial zinc. Furthermore the study of the
resistivity of the film could help us to understand the cause of the
magnetic properties in the films.

By looking at the series of films, we find a tendency: the films are
non-ferromagnetic when grown at 700 ${{}^{\circ }}C$ and are all
ferromagnetic when deposited at 500 ${{}^{\circ }}C$ whatever the oxygen
pressure is. This is evidenced in Figure 3, where the film synthesized at
500 ${{}^{\circ }}C$ displays an hysteresis loop in the ($M-H$) curve with a
saturation value of 1.35 $emu/cm^3$ (corresponding to a saturation
magnetization, $M_S=0.3$ $\mu _{{\bf B}}/Co$) and a coercive field of $%
\approx $100 $Oe$\ (see inset of Fig. 3). A Curie temperature slightly above
300K is also observed on the $M(T)$ curve, confirming the ferromagnetic
behavior.. Below 50 $K$, a rapid increase of the magnetization is also
observed. This paramagnetic contribution might be due to the low
concentration of Co ions in the film which are not coupled to each other.%
\cite{JHKim} On the contrary, the film deposited at a higher temperature (700%
${{}^{\circ }}C$) does not show any hysteresis loop and its magnetization is
close to zero. The situation is more complicated for the films grown at an
intermediate temperature (for example at 600${}$ ${^{\circ }}C$). Indeed,
these films can display a ferromagnetic (as seen in a previous report)\cite
{JAPArno1} or a paramagnetic behavior. However, we have not be able to find
a precise correlation between the observed magnetic behavior and the growth
conditions of the films deposited at 600 ${{}^{\circ }}C$. This indicates
some lack of reproducibility which is more important at this temperature
because 600 ${{}^{\circ }}C$ is an intermediate temperature between the two
regions, and thus a very sensitive one.\cite{NonRepro1,NonRepro2} To
summarize, this study demonstrates first, the importance of temperature
growth on the magnetic properties.\cite{ZnOTemp} Second, the variation of
oxygen pressure at a fixed temperature does not show any influence on the
magnetic measurements: a magnetic transition \thinspace is observed around
300 $K$ for all the films grown at 500${}$ ${^{\circ }}C$. Resistivity
measurements show that the values at 300 $K$ are 0.86 $\Omega .cm$ and 387 $%
\Omega .cm.$ for the film grown at 0.05 Torr and 0.15 $Torr$, respectively.
The rocking curve is constant to a value around 0.43${{}^{\circ }}$, close
to the instrumental limit of 0.15${{}^{\circ }}$ and the one of the Al$_2$O$%
_3$ substrate of 0.2${{}^\circ}$. This indicates that the oxygen pressure
does not have a strong influence on the average crystallinity of the film.%
{\bf \ }The inhomogeneity in the transport measurements compared to the
constant value Curie temperature confirms the independence of the number of
carriers on the magnetic properties, in contradiction with previous reports.%
\cite{DMS1,F} The presence of ferromagnetic or ferrimagnetic Co-based phases
can also be excluded {\it a priori}, due to the low concentration of cobalt
since it is difficult to believe that a small amount of Co will lead to a
magnetic phase that is not observed neither in XRD\ measurements nor in
HRTEM analysis. Furthermore, the shape of the $M(T)$ curve does not exhibit
superparamagnetic behavior with a blocking temperature, but a well defined
Curie transition which means that there is not individual domains of
elements that have ferromagnetic properties.\cite{SuperPara} Anyway, these
results indicate a correlation between the crystallinity of the film and
their magnetic properties.\ Surprisingly, an ill crystallize films (rocking
curve around 0.4-0.5${{}^{\circ }}$) always leads to a ferromagnetic
transition whereas a highly crystallized one (rocking curve around 0.26-0.3${%
{}^{\circ }}$ for the films grown at 700 ${{}^{\circ }}C$) is rarely
ferromagnetic. Thus, we think that a large number of defects is necessary to
observe the ferromagnetic properties which will be discuss later.\cite
{defects}

\subsubsection{Variation with Co content}

Similar kind of study was done keeping the oxygen pressure to 0.1\ Torr
(since it is not influencing the magnetic characteristics) but varying the
Co concentration from 5 to 10 \%, and changing the deposition temperature
between 500 and 700 ${{}^{\circ }}C$. The results show that only the film
grown at 500 ${{}^{\circ }}C$ with 5 \% of Co doping exhibit a (weak)
magnetic signal with a magnetization of 0.1 $emu/cm^3$ (under a magnetic
field of 100 $Oe$) , while the film at 10 \% cobalt is paramagnetic.

In addition, transport measurements revealed that all films are highly
resistive with a value at 300 $K$ of 350 $\Omega .cm$ for the film at 5 \%
and, more than 450 $\Omega .cm$ for the film with 10 \% of Co. This confirms
the previous study\cite{CoZnO2} where the crystallinity was decreasing with
the concentration of cobalt in the film and leads to the creation of
acceptors, increasing the resistivity of the film.

In order to confirm the relations between magnetism and defects, two films
at high (10 \%) and low Co-doping (1.66 \%) were analyzed by TEM (Figure 4).
Fig.4 (a) shows a cross section TEM of a film having 10 \% of Co. At the
interface, a difference of crystallinity between the substrate and the film
is seen. Furthermore, the diffraction pattern of this part shows a well
aligned diffraction peaks assimilate to the substrate with a c-axis
perpendicular to the interface. The film presents different rings, clearly
showing the polycrystalline nature of the structure close to the interface.
Different ED analysis performed on several areas of the film showed that the
crystallinity is increased when approaching the surface. Fig. 4 (b) shows
similar images for a 1.6 \% Co-doped films. Despite a good crystal quality
showed also by ED, some defects can be observed. In the insert of figure 4
(b) is zoomed a tilt of the structure of 60${{}^{\circ }}$. This analysis at
the atomic scale (few nm) must show the presence of any clusters having a
size of minimum 6 nm to be seen by HRTEM.\cite{clusterTEM} This
investigation was carried out on a high number of different crystals for
both films, but no clusters have been detected and cobalt seems to have a
good dilution in the matrix. An analysis with STEM also confirms the good
distribution of the cobalt in ZnO.

\section{Discussion}

Close to the substrate interface, the crystallinity of the film influenced
by the growth temperature and the Co content, is not satisfactory or at
least, not as good as in the upper layers: a\ low substrate temperature
leads to a high disorder of the Co-doped ZnO\ film. This is due to the large
lattice mismatch between the film and the substrate. The effect of the
temperature can be well understood by the increase of kinetic energy of the
particles at the surface of the film, which improves the crystallinity and,
consequently decreases the number of defects.\cite{Revue ZnO}{\bf \ }For
this reason, Saeki {\it et al}. have used a ZnO buffer layer to improve the
quality of the Co-ZnO films. However, in that case a post-annealing was
necessary to obtain a ferromagnetism behavior in the film since in the
as-grown the Co-ZnO is antiferromagnetic.\cite{NonRepro1}

In the present study, it has been showed that in order to get a
ferromagnetic film at 300 $K$, a low Co concentration {\it and} a low growth
temperature are required. If the Co concentration is larger than 5 \%, then
the magnetization becomes negligible. Thus, despite the disorder increasing
in the structure with the amount of cobalt,\cite{JAPArno1} the magnetization
shows a paramagnetic behavior.

Furthermore, the formation of any clusters or second phases was {\it a priori%
} minimized because of the particular deposition technique which is
alternative deposition. This process indeed favors the dispersion of the
cobalt inside the structure. For this, two metallic targets are held on a
carrousel. The laser, using the rotation of the two targets, fired on the
zinc followed by the cobalt target. For example, to obtain 5 \% cobalt doped
film, it is necessary to fire 19 pulses on Zn target and then 1 pulse on the
Co target. The deposition rate is close to 0.052 $nm/pulse$. Thus, one
cobalt pulse is sandwiched between 2 ZnO layers made up 19 pulses each. It
is unlikely that the quantity of cobalt between two layers of ZnO is
sufficient to create some cobalt clusters. Furthermore, a gradient of
diffusion (due to the temperature of the substrate) will appear, favoring
the migration and the dilution of cobalt in the ZnO structure. A low growth
temperature unlikely decreases the Co diffusion, and might induce the
presence of Co-rich phases (i.e. Co clusters or secondary phases). These
phases can lead to ferromagnetism or ferrimagnetism behaviors.\ Such result
might explain the similarity between the films grown at 500${{}^{\circ }}C$
under various oxygen pressure. However, based on the XRD and the HRTEM no
Co-rich phases have been evidenced and, this model can not explain the
antiferromagnetic behavior at high Co concentration.

Another possible explanation for the observation of ferromagnetic is as
follow, depending on the crystallinity and the Co concentration. At low
doping, the short interactions mechanisms like superexchange and Zener
double-exchange interaction can be avoided, because of the low probability
to have two cobalt neighbor atoms. At longer range ferromagnetic exchange,
the $RKKY$ model can be also avoided due to the $n$-type behavior of ZnO and
the low number of carriers in the films. Thus, these models can not be used
to explain the ferromagnetism. We believe that the recent model of the bound
magnetic polaron used in the $n$-type semiconductors by shallow donors is
more appropriate.\cite{Int1} In this model, Coey {\it et al}. utilized the
general formula for the oxide: $(A_{1-x}Co_x)(O\square _\partial )$ where $A$
is nonmagnetic cation and $\partial $ is the donor defect.\cite{Int1}

Experimentally, this means that the ferromagnetism appears only with a high
probability of donor defects in the film, at low deposition temperature
compared to 700${}$ ${^{\circ }}C$ where the crystal defects are less
present. Finally, the high defect concentration close to the interface leads
to the formation of an impurity band, which is polarized by exchange with
magnetic elements. This is confirmed by the value of the saturation
magnetization which is equal to ($0.3\pm 0.1)$ $\mu _B/Co.$ Surprisingly,
this value is smaller compared to the Co${{}^2}$$^{+}$ spin state in a
tetrahedral crystal field (low spin= $1$ $\mu _B$ or high spin= $3$ $\mu _B$%
). However, if we only consider the cobalt atoms having an environment of
defects that are at the origin of ferromagnetism (i.e. for example, $1/3$ of
the film), the value would be in the order of $1$ $\mu _B/Co$. This
reinforced the fact that the ferromagnetism originated from the ''bulk''
material and this is likely the origin of ferromagnetism at low doping
concentration.\cite{Bulk} It should also be pointed out that in these
experiments, the oxygen pressure induces large changes in resistivity (see
above). As consequence and, based on this model, the oxygen pressure should
have an effect on the magnetization.\ Since it is not the case, this means
that a detailed study on the nature of the defects is required (structural
defects, oxygen and zinc vacancies, interstitials etc.).\cite{defectsa}

At high doping, the situation is more simple.\ Due to the antiferromagnetic
coupling between the Co atoms, which comes from the short magnetic
interactions (resulting from the high probability to have two cobalt
neighbors), the sample exhibits paramagnetism.

\section{Conclusion}

The comparison of microstructural and macrostructural analysis of a series
of Co-doped ZnO films clarified the conditions to obtain interesting
magnetic properties. Only the films at low cobalt doping {\it and} grown at
low temperature display ferromagnetism at room temperature. Whereas the
number of carrier in the film does not seem to control the magnetism, the
defects are necessary to observe ferromagnetic properties. In spite of the
high concentration of cobalt in the structure, the microstructural analysis
do not show cobalt clusters, revealing an homogeneous cobalt distribution in
the ZnO matrix. This is confirmed by the magnetic measurement which do not
exhibiting superparamagnetism behavior. Based on these experiments, we
believe that the ferromagnetism is intrinsic and result from long-range
interactions induced by the defects. The determination of synthesis
condition favoring the reproducibility of magnetic properties might be the
beginning point from the elaboration of some devices for the spin
polarization.

\smallskip

Partial financial support from the Center Franco-Indien pour la promotion de
la recherche Avanc\'{e}e/Indo-French Center for the promotion of advanced
Research (CEFIPRA/IFCPAR) under Project No. 2808-1 is acknowledged. The
Centre National de la Recherche Scientifique (CNRS) and the ''Conseil
R\'{e}gional de Basse Normandie'' is also supporting this work through a BDI
fellowship. 
\newpage

Figures Captions:

Figure 1: High resolution transmission electron microscopic image of ZnO
film and Fourier transformation pattern (inset). All crystal fragments are
oriented with c-axis parallel to the electron beam. (a) Film grown at 600 ${%
{}^{\circ }}C$ with on the Fourier transformation, 6 peaks characteristic of
the 6-fold symmetry of ZnO. The pattern (b) and (c) are film grown at T=700 $%
{{}^{\circ }}C.$ The first one shows HRTEM\ of crystal in film and the last
one shows the interface with ''moir\'{e}'' phenomena.

Figure 2: Different diffraction pattern taken from the interface of the film
until close to the surface. An evolution of the crystallinity is observed
with 6 well defined peaks close to the surface and a broadening of these
diffraction peaks along the $c$-axis until the interface.

Figure 3: (M-T) curve of zero field cooled and field cooled (100 Oe) for a
Co:ZnO: 1.66 \% thin film grown at 500 ${{}^{\circ }}C$. The inset presents
the (M-H) curves at 10K of films grown at 500 and 700 ${{}^{\circ }}C.$ The
coercive field of the film grown at 500 ${{}^{\circ }}C$ is about 100 $Oe$.
All these curves have corrected from the substrate component.

Figure 4: HRTEM images of Co-ZnO\ films deposited at 600 ${{}^{\circ }}C.$
(a) is a cross section of a high doping Co-ZnO: 10 \%. The well crystallized
right part is the substrate and at left the film. The inset is the
diffraction pattern of the interface. The aligned peaks are the substrate
and the rings are due to the film. (b) is the film doped with 1.6 \% cobalt
with $a$ and $b$ axis in-plane.\newpage


\begin{thebibliography}{00}
\bibitem{Bulk}  In that case, ''bulk'' is associated to intrinsic
phenomena.\ It means the overall ''bulk'' film by opposition to the clusters
or secondary phase present in a film.
\end{thebibliography}
\end{document}